# Shear Induced Demixing in Bidisperse and Polydisperse Polymer Blends: Predictions From a Multi-Fluid Model

Joseph D. Peterson, Glenn H. Fredrickson, L. Gary Leal

*Department of Chemical Engineering, Santa Barbara, CA 93107*

## Abstract

In light of recent advancements in the constitutive modelling of bi-disperse and polydisperse entangled linear polymers, we present a new 'multi-fluid' generalization of the classic two-fluid approximation for flows of inhomogeneous polymer blends. As an application of the model, we consider predictions for the linear and nonlinear dynamics of shear induced demixing (SID) instabilities in blends with bi-disperse and log-normal molecular weight distributions. We find that even in the absence of any chemical contrast between component chains, an imposed flow can induce a demixing instability provided there is sufficient contrast in the size of the two chains. The lower bound polydispersity for SID coincides with the point where elastic forces ($k_B T$ per entanglement) scaled by the contrast between chains (e.g. polydispersity index minus one) exceed the entropic forces for mixing ($k_B T$ per chain). For bi-disperse blends, we show that the non-linear dynamics of SID strongly resemble what has previously been shown for SID in entangled polymer solutions.

## Section I: Introduction

More than half a century ago, it was found that highly entangled polymer melts can 'fractionate' in flow, leading to a spatially inhomogeneous molecular weight distribution [1] [2] [3] [4]. Specifically, for capillary flows of high molecular weight polyethylene ($M_w > 10^5$g/mol), experiments have shown that long chains tend to migrate towards the centerline and short chains tend to migrate towards the walls [2] [3] [4]. At the same time, however, polystyrene melts of comparable molecular weight do not appear to fractionate in flow [5]. Historically, this discrepancy has been explained in terms of differing tendencies for crystallization [4], but in the present paper, we suggest that a difference in the entanglement molecular weights may be more important.

More dramatic and unambiguous evidence of flow-induced fractionation has been found for well-entangled mixtures of miscible but chemically dissimilar polymers. For example, mixtures of polystyrene (PS) and polyvinylmethylether (PVME) exhibit an experimentally accessible lower critical solution temperature, and the effective miscibility of the components can be altered by an imposed flow. Depending on the applied shear rate, both shear induced demixing (low shear rates) and shear induced mixing (higher shear rates) have been observed [5] [6] [7].

The earliest theoretical explanations of flow-induced fractionation phenomena in capillary flows were built on a quasi-thermodynamic argument; long chains migrate away from strong shear to minimize an elastic contribution to the system's free energy [1] [3]. These proposals lack the rigor and theoretical foundations of modern two-fluid models [8] [9] [10], but were nonetheless insightful for establishing a connection between elastic forces and polymer migration. Quasi-thermodynamic descriptions of flow induced fractionation have since been supplanted by a more general 'two-fluid' framework [8] [11]. In a two-fluid model, the components of the blend are envisioned as a pair of superimposable continuum fluids whose relative/collective motions are coupled by friction and incompressibility. Polymer migration across streamlines (and hence fractionation) arises via the requirements of force balance; whenever stresses are unequally divided between the components, the imbalance of forces acting on individual chains drives a relative migration. In a polymer solution, this migration can amplify thermal fluctuations in concentration [12] [13] and also lead to a compositionally inhomogeneous flow [14]. In polymer melts, a similar set of physics is expected to produce spatial gradients in the molecular weight distribution [8].

Two-fluid models of polymer blends were first introduced by Doi and Onuki [8], and the same framework has since been applied in several studies of flow induced mixing and demixing in simple shear flows of bimodal polymer blends [9] [15] [10] [16] [11]. Because migration of polymers can be driven by stresses incurred during flow, predictions for shear induced mixing and demixing are sensitive to the details of the constitutive model that one employs [9]. Early work by Clarke and McLeish used a Doi-Edwards constitutive model to show that shear flow can favor either mixing or demixing depending on both the blend composition and the applied flow rate. However, this work also explicitly neglected any isotropic elastic contribution to the stress – while this may be appropriate for modelling incompressible homogeneous flows, it is not appropriate for two-fluid systems since each component fluid is individually compressible. Later work by Jupp and Yuan employed bi-disperse generalizations of the Johnson-Segalman and Rolie-Poly fluids, but these models are also problematic: neither one describes the effect of polydispersity on stretch relaxation dynamics [17] [18], for example. Worse yet, the Rolie-Poly instance is not self-consistent when the bidisperse blend is a mixture of identical chains.

Recent developments in the non-linear constitutive modelling of bi-disperse and polydisperse linear-chain blends [19] have created an opportunity to revisit the problem of inhomogeneous polymer blends in flow. Here, we present a generalized 'multi-fluid' framework suitable for studying inhomogeneous flows of bi-disperse and polydisperse polymer blends. As a first application of the model, we present results on the linear and nonlinear dynamics of SID instabilities in blends of chemically homogeneous linear-chain polymers. A chemically homogeneous polymer blend is a blend of polymers with differing molecular weights but a shared monomer basis. Compared to polymer solutions, polymer blends have more applications in industry and yet they have received less attention in the experimental and theoretical literature for SID.

With the new multi-fluid model in place, we revisit the problem of SID in bi-disperse blends. In our study, we find that the lower bound polydispersity for SID coincides with the point where elastic forces ($k_B T$ per entanglement) scaled by the typical contrast between chains (e.g. polydispersity index minus one) exceed the entropic forces for mixing ($k_B T$ per chain)[1]. With regard to the dynamics of SID, we observe the growth of a 'layered' structure in the flow gradient direction with the melt composition (and shear rate) alternating between layers. Predictably, the composition and shear rate contrast between the layers depends on the molecular weight contrast in the component polymers. In the early stages of SID, the typical layer thickness is comparable to the typical polymer coil size, but over longer timescales the layers are predicted to merge and form bands of macroscopic dimension. These observations are consistent with previously known predictions for the dynamics of SID in polymer solutions [20].

Finally, we present results from a study of polydisperse blends with log-normal molecular weight distributions. These results confirm that the preceding predictions of SID are not a peculiarity of bi-disperse blends, but rather a feature of polydisperse blends in general.

## Section II: Governing Equations
### Section II.A: Discretizing a continuous molecular weight distribution

We consider an arbitrary molecular weight distribution $\phi(Z)$, defined such that the volume fraction of chains with entanglement numbers $Z \in [Z, Z + \delta Z]$ is given by $\phi(Z)\delta Z$. To construct a multi-fluid model for the inhomogeneous flow of such a melt, we first must transform the continuous molecular weight distribution into a set of discrete fractions to be treated as independent fluids. As with any discretization process, one must confirm that the predictions of the multi-fluid model converge with increasing resolution of the molecular weight distribution.

For bi-disperse blends, the molecular weight distribution is already properly discrete – we denote the long/short chain entanglement numbers as $Z_L/Z_S$, and their respective volume fractions as $\phi_L/\phi_S$. For polydisperse melts, converged predictions should be independent of the discretization scheme provided the discretization scheme converges to the correct cumulative distribution function. An example of such a discretization scheme is outlined in the work by Boudara et. al. [19]. However, this discretization scheme requires that one truncate the tails of the molecular weight distribution at some point – this is easily done for a single molecular weight distribution, but more difficult to do in a consistent way when studying and comparing results across a range of polydispersities. Therefore, we have found it more convenient to employ an 'adaptive' discretization scheme in which the molecular weight distribution is partitioned into sections of equal volume fraction with each section represented by its weight average molecular weight. We index the entanglement numbers and volume fractions of the $N$ discrete fractions by $Z_1, Z_2, Z_3 \ldots Z_N$ and $\phi_1, \phi_2, \phi_3, \ldots \phi_N$, respectively[2]. From this point forward, the $N$ fractions

---

[1] Although our study does not explicitly consider shear induced mixing (SIM) in non-homologous blends, one can levy similar scaling arguments to suggest that shear flow alone is likely a poor tool for compatibilizing immiscible blends. For immiscible blends, the elastic forces favoring mixing are at most $k_B T$ per entanglement, while the driving forces for segregation scale as $k_B T$ per monomer.

[2] For a system with $M$ distinct chemistries in the blend, one must discretize the molecular weight distribution for each chemistry independently. Using $N_1, N_2, N_3, \ldots N_M$ fractions for each component one obtains a corresponding set of volume fractions $\phi_1, \phi_2, \ldots \phi_{N_1+1}, \phi_{N_1+2}, \ldots \phi_{N_1+N_2+1}, \ldots \phi_{N_1+N_2+\cdots N_M}$ and entanglement

of the blend will be treated as a set of independent and superimposable continuum fluid with velocities $u_1, u_2, u_3, \ldots u_N$.

## Section II.B: Dimensional equations (tensorial)

A natural outcome of the multi-fluid framework is that the total stress in the melt can be unequally divided amongst its components. When this happens, the imbalance of elastic stresses drives polymers to migrate, and the migration is resisted only by osmotic forces favoring mixing and by a frictional drag force against neighboring constraints (i.e. against its confining tube that moves with velocity $u_T$). A force balance equation summarizing these effects can be produced using the Rayleighian formalism of Doi and Onuki, from which we find that the velocity of fluid component $i$, $u_i$, differs from the tube velocity, $u_T$ [21] by:

$$\zeta_i (u_i - u_T) = \nabla \frac{\delta F}{\delta \phi_i} + \nabla P - \frac{\zeta_i}{\bar{\zeta}} \mathcal{F} - \eta_D \nabla^2 \langle u \rangle \qquad (1)$$

Here $\zeta_i$ is the drag coefficient per unit volume of species $i$ with its confining tube, $\bar{\zeta}$ is the mean drag coefficient (c.f. equation (2)) $F$ is the system free energy, $P$ is the total pressure, and $\eta_D$ is a viscosity for any small Newtonian contribution to the stress associated with the volume-averaged velocity of the entire blend, $\langle u \rangle = \sum_i \phi_i u_i$. $\mathcal{F}$ represents a set of elastic body forces per unit volume acting on species $i$ due to the stress being unequally divided amongst the components.

Equation (1) shows that the friction generated by a chain moving against its constraints (left hand side) is matched by (right hand side) driving forces from chemical potential gradients (first term) pressure gradients (second term), purely elastic stresses (third term) and purely viscous stresses (final term). By performing a force balance on the tube itself, we can define the tube velocity as a frictionally-weighted average velocity of the component species [21]:

$$u_T = \frac{1}{\bar{\zeta}} \sum_m \zeta_m \phi_m u_m \qquad \bar{\zeta} = \sum_m \zeta_m \phi_m \qquad (2)$$

The system free energy $F$ includes both elastic and osmotic contributions. For the osmotic contributions, we use a Flory Huggins model and a squared gradient potential for the free energy of mixing $F^{mix}$:

$$F^{mix} = \frac{k_B T}{b^3} \int dV \sum_{i=1}^{N} \left[ \frac{\phi_i}{Z_i N_e} \ln(\phi_i) + \frac{b^2}{36 \phi_i} |\nabla \phi_i|^2 + \frac{1}{2} \sum_{j=1}^{N} \phi_i \phi_j \chi_{ij} \right] \qquad (3)$$

where $b$ is the Kuhn monomer size, $N_e$ is the number of Kuhn monomers per entanglement, and $\chi_{ij}$ is the Flory interaction parameter (in units of $k_B T / b^3$) for species $i$ and $j$. Through most of the present

---

numbers $Z_1, Z_2, \ldots Z_{N_1+1}, Z_{N_1+2}, \ldots Z_{N_1+N_2+1}, \ldots Z_{N_1+N_2+\cdots N_M}$. The governing equations of the multi-fluid model follow in the same way, with the total number of fluids $N$ given by $N = N_1 + N_2 + \cdots N_M$.

paper, we are primarily concerned with melts of chemically identical polymers ($\chi_{ij} = 0$) but we will provide some discussion on the effect of a chemical contrast.

To this point, the basic structure of our multi-fluid model is in agreement with other two-fluid models for polymer blends. Moving forward, however, we offer two new improvements. First, we present a slightly different view for how stresses arise from an elastic contribution to the free energy. Second, and more importantly, we link our momentum balance equations to an improved constitutive model.

In most prior two-fluid model calculations, the free energy $F$ contains no explicit terms relating to the elastic free energy. Instead, the elastic free only appears implicitly (via the stress, which is assumed to be elastic in nature) when estimating the virtual work done by a small deformation. With this approach, the expression for elastic forces driving migration $\boldsymbol{\mathcal{F}}_{DO}$ (subscript denotes an attribution to Doi and Onuki [8]) is given by:

$$\boldsymbol{\mathcal{F}}_{DO} = \nabla \cdot \boldsymbol{\sigma} \qquad (4)$$

where $\boldsymbol{\sigma}$ is the total elastic stress tensor for the whole melt. However, a slightly different result can be obtained if one admits explicit elastic terms into the free energy. On timescales comparable to the 'entanglement time' (where a single entanglement strand explores its available configurations) one can ignore stress relaxation dynamics and invoke a 'local equilibrium' approximation to describe the stresses within the polymer using rubber elasticity theory. Treating entanglements as temporary cross-links, we assume that the total elastic free energy of the system $F^{el}$ is given by a summation over the elastic free energy held by each individual point of entanglement between two polymers:

$$F^{el} = \int dV \left[ \sum_i \sum_j \phi_i \phi_j f_{ij}^{el} \right] \qquad (5)$$

$$f_{ij}^{el} = \frac{1}{2} G \left( \text{trace}(\boldsymbol{Q}_{ij}) - 3 - \ln(\det(\boldsymbol{Q}_{ij})) \right) \qquad (6)$$

where $G$ is the shear modulus and $\boldsymbol{Q}_{ij}$ is a conformation tensor that describes the effect of entanglement between chains from fluids $i$ and $j$. For bi-disperse blends, Daniel Read has shown that our expression for the elastic free energy can be derived from a tube-based model in the particular case of bi-disperse blends – unfortunately, that derivation is not yet available for publication but should be soon. In the meantime, the reader can interpret equation (5) as a reasonable first guess that describes the whole entangled structure as a superposition of cross-linked gels, where each gel represents a different subset of entanglement types. With this explicit form for the elastic free energy, we find that $\boldsymbol{\mathcal{F}}$ is given by:

$$\boldsymbol{\mathcal{F}} = \nabla \cdot \boldsymbol{\sigma} + \sum_i \sum_j \nabla \boldsymbol{Q}_{ij} : \frac{\delta F}{\delta \boldsymbol{Q}_{ij}} \qquad (7)$$

$$\sigma = \sum_i \sum_j 2\boldsymbol{Q}_{ij} \cdot \frac{\delta F}{\delta \boldsymbol{Q}_{ij}} = G \sum_i \sum_j \phi_i \phi_j (\boldsymbol{Q}_{ij} - \boldsymbol{I}) \qquad (8)$$

$$\nabla \boldsymbol{Q}_{ij} : \frac{\delta F}{\delta \boldsymbol{Q}_{ij}} = \frac{1}{2} \phi_i \phi_j \nabla f_{ij}^{el} \qquad (9)$$

Here, once again, $\sigma$ is the 'total' stress tensor in the sense that it contains all of the elastic stresses that one might measure from bulk rheology, so equation (8) also defines the stress tensor used in equation (4). Thus, admitting an explicit form to the elastic free energy introduces additional terms into the force balance equation (1) via $\mathcal{F}$ and the chemical potential $\delta F/\delta \phi_i$[3]. However, there is a potential conflict between the 'local equilibrium' approximation invoked for deriving (5) and the virtual work calculations employed when deriving (7). As such, it is not yet a-priori obvious whether the implicit or explicit treatment of the elastic free energy is most appropriate, and so this paper will provide calculations from both frameworks.

Our next task is to specify a constitutive model through an equation of motion for $\boldsymbol{Q}_{ij}$. The evolution of each $\boldsymbol{Q}_{ij}$ tensor is assumed to follow the 'Symmetric Rolie-Double-Poly' model (SRDP) [19], which is an amalgam of the double reptation ansatz [22] and the Rolie-Poly model [23]. The SRDP model describes stress relaxation by reptation, $\boldsymbol{R}_{rep,ij}$, chain retraction, $\boldsymbol{R}_{ret,ij}$ and convective constraint release (CCR), $\boldsymbol{R}_{CCR,ij}$:

$$\overset{\nabla}{\boldsymbol{Q}}_{ij} = -\boldsymbol{R}_{rep,ij} - \boldsymbol{R}_{ret,ij} - \boldsymbol{R}_{CCR,ij} \qquad (10)$$

$$\boldsymbol{R}_{rep,ij} = \frac{1}{2} \left[ \frac{1}{\tau_{D,i}} + \frac{1}{\tau_{D,j}} \right] (\boldsymbol{Q}_{ij} - \boldsymbol{I}) \qquad (11)$$

$$\boldsymbol{R}_{ret,ij} = \frac{1}{2} [f_{ret,i} + f_{ret,j}] \boldsymbol{Q}_{ij} \qquad (12)$$

$$f_{ret,i} = \frac{2}{\tau_{R,i}} \left( 1 - \frac{1}{\lambda_i} \right) \qquad \lambda_i = \left[ \frac{1}{3} \sum_j \phi_j \text{tr}(\boldsymbol{Q}_{ij}) \right]^{1/2} \qquad (13)$$

$$\boldsymbol{R}_{CCR,ij} = \frac{1}{2} \left[ \frac{f_{ret,i}}{\lambda_i} + \frac{f_{ret,j}}{\lambda_j} \right] (\boldsymbol{Q}_{ij} - \boldsymbol{I}) \qquad (14)$$

where $\tau_{D,i} = \tau_D(Z_i) = 3Z_i^3 \tau_e$ and $\tau_{R,i} = \tau_R(Z_i) = Z_i^2 \tau_e$ are the reptation and longest Rouse time, respectively, for a monodisperse melt of chains with entanglement number $Z_i$ based on the Rouse time of an entanglement segment, $\tau_e$. Further corrections to the relaxation spectra accounting for the effects of a polydisperse environment are described in our recent paper [19] but are omitted here for simplicity. For multi-fluid model calculations, the upper convected Maxwell derivative must employ the 'tube' velocity, as it is the motion of the confining tube that is ultimately responsible for deformation of polymers in flow:

$$\overset{\triangledown}{Q}_{ij} = \frac{\partial Q_{ij}}{\partial t} + u_T \cdot \nabla Q_{ij} - (\nabla u_T)^T \cdot Q_{ij} - Q_{ij} \cdot \nabla u_T \qquad (15)$$

---

[3] A similar issue arises in polymer solution models, where an elastic osmotic pressure term appears if an elastic contribution to the free energy is explicitly included [20].

Finally, the migration of polymers across streamlines will lead to changes in the composition:

$$\frac{\partial}{\partial t}\phi_i = -\nabla \cdot (\phi_i u_i) \qquad (16)$$

Summing equation (16) over all species yields the usual constraint for incompressibility of the overall volume averaged flow, $\nabla \cdot \langle u \rangle = 0$.

There is one additional note on the drag coefficients, $\zeta_i$. Each $u_i$ represents a center-of-mass velocity, but drag against the tube must consider the curvilinear velocity of a polymer within its tube: therefore, the drag coefficients $\zeta_i$ must scale with the chain's contour length, $\zeta_i = \zeta_0 Z_i$ [21]. For chains that are nearly disentangled, $Z_m = 1$, a Rouse model (neglecting entanglements) prescribes a drag coefficient $\zeta = \zeta_0 = \mu_e/a^2$, where $\mu_e$ is the viscosity of a melt at its entanglement weight and $a = bN_e^{1/2}$ is the tube diameter (radius of gyration for a chain at its entanglement weight) [24].

To close the above equations, one will require (1) initial conditions on the polymer configuration tensors $Q_{ij}$ and volume fractions $\phi_i$ and (2) boundary conditions on the component velocity fields $u_i$ and volume fractions $\phi_i$. At any solid boundary moving with velocity $u_W$ one can enforce no slip and no flux boundary conditions on the velocity field components, $u_i = u_w$. One also requires boundary conditions on each component volume fraction: if a solid boundary is neutrally interacting with no chemical affinity for one species over another, then a zero gradient boundary conditions is appropriate $n \cdot \nabla \phi_i = 0$. No boundary conditions on the configuration tensors $Q_{ij}$ are needed at a solid boundary.

# Section III: Simple Shear Flow
## Section III.A: Discussion of Dimensionless Groups

For the remainder of the present paper, we will be applying the equations outlined in the preceding section to study '1D' simple shear flows of bidisperse and polydisperse polymer blends.

The '1D' simple shear flow geometry is defined by a pair of infinite parallel plates separated by a distance $H$ in the $y$-direction, with an $N$-component SRDP blend filling the gap in between. The top plate is translated in the $x$-direction at a velocity $U$ relative to the bottom plate, and the flow/composition is assumed to be uniform in the directions of flow ($x$-direction) and vorticity ($z$-direction), varying only in one dimension ('1D'), namely the velocity gradient direction ($y$-direction).

This '1D' approximation reduces the dimensionality of the problem while preserving physics pertinent to the linear and non-linear dynamics of a shear induced demixing instability. With regards to the linear dynamics of SID, prior studies of SID in viscoelastic two-fluid models have shown that linear modes of instability are plane-waves oriented in the flow gradient direction [12] [13] [25], for which our 1D approximation is adequate. With regards to the non-linear dynamics of SID, prior studies of a viscoelastic two-fluid model in 2D have shown that SID leads to a flow that is stratified (or 'banded') in the flow gradient direction and slowly coarsens overtime [11] [16] [10]. Broadly speaking, this is in good agreement with what the 1D approximation predicts, albeit with differing details of the coarsening mechanism [14] [26]. The aforementioned 2D studies were nominally designed to consider SID in polymer blends, but whereas we trust the resulting predictions as a general representation of SID in a viscoelastic two-fluid model, we do not trust that the parameter space considered by the authors (or the details of the constitutive model) are suitable for describing polymer blends in particular.

The full dimensionless governing equations of relevance to the present study are outlined in the appendix. Here, we will review and discuss the dimensionless groups that govern the linear and non-linear dynamics of SID in 1D, and afterwards we will compare these dimensionless groups to the ones that emerge in the study of SID in polymer solutions.

For a given molecular weight distribution, we find that the dynamics of SID in polymer blends are governed by three dimensionless groups:

First is the Weissenberg number, $Wi = U\tau_D(\bar{Z})/H$, which is a characteristic rate of strain relative to the rate of stress relaxation for a chain with $\bar{Z} = \sum_i \phi_i Z_i$ entanglements per chain. For the special case of bi-disperse blends, a longest relaxation time is more well-defined and we find it more convenient to define the Weissenberg number with respect to the long chain reptation time instead, $Wi_L = U\tau_D(Z_L)/H$.

Second is the dimensionless radius of gyration, $\bar{R}_g = b(N_e\bar{Z})^{1/2}/H$, which compares the typical coil size to the gap dimension. This parameter sets both the minimum length-scale for changes in material composition and the characteristic length for which diffusion and stress relaxation happen at the same rate.

Third is the ratio $\eta = \eta_D/G\tau_D(\bar{Z})$, which compares the importance of purely viscous and purely elastic stresses in the problem. If we suppose that the purely viscous forces are no larger than the Rouse viscosity of an entanglement segment $\eta_D \sim \mu_e$, then $\eta \sim \mathcal{O}(\bar{Z}^{-3}) \ll 1$. From our prior work on entangled polymer solutions in simple shear flow, we found that viscous forces can often be ignored when they are small compared to elastic stresses. Therefore, to simplify the analysis of this report, we primarily focus on the limiting case of $\eta = 0$, but we remind the reader that this simplification is unique to 1D simple shear flow with neutrally interacting solid boundaries – in general, one cannot enforce $\boldsymbol{u}_i = \boldsymbol{u}_W$ for all components at a solid boundary without some finite value of $\eta$.

Finally, when the molecular weight distribution is not considered fixed, any independent moments of the molecular weight distribution $\phi(Z)$ can be considered dimensionless groups:

First, the mean entanglement number, $\bar{Z} = \sum \phi_i Z_i$, is the average number of entanglements per chain across the entire system. The mean entanglement number controls the relative strength of elastic and osmotic forces involved in polymer migration, as well as the typical ratio of time-scales for reptation and chain retraction.

Second, the polydispersity index, $I_P = \bar{Z}\sum\phi_i/Z_i$ is a measure of the typical size contrast between chains in the blend. For $I_P - 1 \ll 1$, the chains are all nearly identical, and for $I_P \gg 1$ there is considerable variation to consider. The polydispersity index controls the rheological contrast between chains, and therefore the degree to which elastic forces are able to effect migration.

Third, one might consider skewness of the molecular weight distribution. Note, however, that for log-normal molecular weight distributions, the skewness is not independent of the polydispersity. For a bi-disperse blend of given $\bar{Z}$ and $I_P$, the skewness of the distribution can be described in terms of the mean fraction of long chains, $\phi_L$.

In the case of a multicomponent mixture with $M$ chemically distinct species, there will be an additional $M(M-1)/2$ dimensionless groups related to the set of Flory interaction parameters, $N_e\chi_{ij}$, that describe the scale of enthalpic mixing forces relative to elastic forces. These are typically not considered in the present study, where we are primarily focused on blends of chemically identical chains, $M = 1$.

Absent from the list is any direct dependence on the average molecular weight of polymers in a melt – the molecular weight only appears indirectly via the entanglement number $\bar{Z}$. As such, we might expect that two blends with similar molecular weight distributions will show different tendencies to fractionate or demix in flow if they have very different entanglement molecular weights. This is in qualitative agreement with experiments, which have shown that polyethylene melts are far more susceptible to fractionation in capillary flows than polystyrene melts of comparable molecular weight [2] [3] [4] [5]. Note that the entanglement weights of these two polymers differ by roughly a factor of twenty [24]. Previous studies have explained the disparity in terms of differing tendencies to crystallization [4], and additional experiments would be needed to determine which of these proposed mechanisms is most important.

In summary, besides the dimensionless terms that describe the molecular weight distribution ($\bar{Z}, I_P, \phi_L$), there are only two dimensionless groups relevant to the problem: $Wi$ and $\bar{R}_g$ (here we ignore $\eta$ on the basis of $\eta \ll 1$). However, when the gap dimension is large relative to the typical coil size ($\bar{R}_g \ll 1$), the stability criterion for SID is (to leading order) independent of $\bar{R}_g$ and depends only on the value of $Wi$. What remains to be understood are the details for how the molecular weight distribution contributes to mixing or demixing at a given $Wi$.

For any reader familiar with the existing literature on SID in polymer solutions, we provide an appendix to discuss the relationship between our blend model and a corresponding polymer solution model [14] [27] [20].

## Section IV: Calculations for bi-disperse blends

Bidisperse blends are the most elementary class of blends and therefore present a natural starting point for our investigation of SID. In this section, we will present a comprehensive study for how the composition of a blend ($\bar{Z}, I_P, \phi_L$) affects its stability to SID over a wide range of Weissenberg numbers. A principal finding of this section is that the lower bound polydispersity for SID coincides with the point where elastic forces ($k_BT$ per entanglement) scaled by the contrast between chains ($I_P - 1$) exceed the entropic forces for mixing ($k_B T$ per chain), or $I_P - 1 \sim \mathcal{O}(1/\bar{Z})$. For very well entangled polymer blends, this leads to a surprisingly strict upper bound on what may be considered 'monodisperse' from the standpoint of SID: for example, a blend with $\bar{Z} = 100$ entanglements per chain may be susceptible to SID if the polydispersity exceeds $I_P \sim 1.10$.

## Section IV.A: Linear Stability: Long wavelengths

A compositionally homogeneous flow is a solution to the steady-state governing equations, but it is not necessarily a stable solution. To consider a system's response to infinitesimal perturbations about a steady homogeneous flow, we separate all variables into their steady solutions plus a small perturbation that evolves in time. Denoting the steady solutions to a variable $X$ as $X^0$ and perturbations by $\delta X$, we have $\phi_i = \phi_i^0 + \delta \phi_i$, $Q_{ij} = Q_{ij}^0 + \delta Q_{ij}$, $u_i = u_i^0 + \delta u_i$ and so on. We insert these expressions into the governing equations and (since the perturbation is infinitesimal) retain only the linear perturbation terms. Because we perform our linear stability analysis on a homogeneous base state, the spatial dependence of the eigenmodes will be strictly sinusoidal, and so each perturbation $\delta X$ we be represented by $\delta X(t,y) = \delta X^0(t) \exp(iky)$. In our non-dimensionalization scheme, the wavenumber $k$ is given in units of $1/H$.

What results is a linear system of differential and algebraic equations that is first-order in time. The algebraic equations (i.e. the force balance and incompressibility relations) slave changes in velocity to changes in concentration and configuration, e.g. $\delta u_i^0 = \sum_j \alpha_{ij} \delta \phi_j^0 + \sum_k \sum_j \sum_l \sum_m \beta_{ijklm} \delta Q_{jk,lm}^0$. Following this simplification, what remains is a linear ODE matrix equation:

$$\partial_t \delta \boldsymbol{X}^0 = \boldsymbol{M}(k\bar{R}_g, \bar{Z}, I_P, \phi_L, Wi) \delta \boldsymbol{X}^0 \qquad (17)$$

where $\delta \boldsymbol{X}^0 = \{\delta\phi_1, \delta\phi_2, \ldots \delta\phi_{N-1}, \delta Q_{11,xx}, \delta Q_{11,yy}, \delta Q_{11,zz} \ldots \delta Q_{NN,zz}\}$ is a vector of all linearly independent perturbation variables to which a time derivative is applied in the governing equations. The eigenvectors/eigenvalues of the matrix $\boldsymbol{M}$ provide the perturbations/growth rates (in units of $1/\tau_D(\bar{Z})$) that exhibit single-exponential growth/decay. We denote $\sigma$ as the real component of the largest eigenvalue in $\boldsymbol{M}$, and whenever $\sigma > 0$, homogeneous flow is linearly unstable to SID.

In general, the stability to SID depends on the wavenumber of the perturbation under consideration. However, when a typical polymer coil is much smaller than the gap size, $\bar{R}_g \ll 1$, as is typically the case in an experimental setting, the absolute stability to SID is virtually independent of the wavenumber for long wavelength perturbations, $k\bar{R}_g \ll 1$. The same is true for SID in polymer solutions [27]. Therefore, to consider the absolute stability to SID in a real experimental setting, we need only describe stability boundaries for long wavelength perturbations, $k\bar{R}_g \ll 1$.

For long wavelength perturbations, $k\bar{R}_g \ll 1$, growth or decay is limited by diffusion (as opposed to stress relaxation) and the leading order wave-number dependence of the growth rate goes as $\sigma \sim \sigma_0 (k\bar{R}_g)^2$. The absolute stability to SID is then determined by the sign of $\sigma_0$. Additional details for evaluating linear stability in the $k\bar{R}_g \ll 1$ limit are provided in the appendix.

In the figures that follow, we present the neutral stability boundary, $\sigma_0 = 0$, for macroscopic systems $k\bar{R}_g \ll 1$ over a wide range of bidisperse blend compositions and flow rates. First, for a blend with $\phi_L = 0.5$ we fix the long chain entanglement number to $Z_L = 30, 100, 300, 1000$ and vary the short chain entanglement number $Z_S$ over the range $2 < Z_S < Z_L$. We also vary the Weissenberg number over the range $10^{-2} < Wi_L < 10^2$. For each blend composition, we plot a curve representing the neutral stability boundary to shear induced demixing.

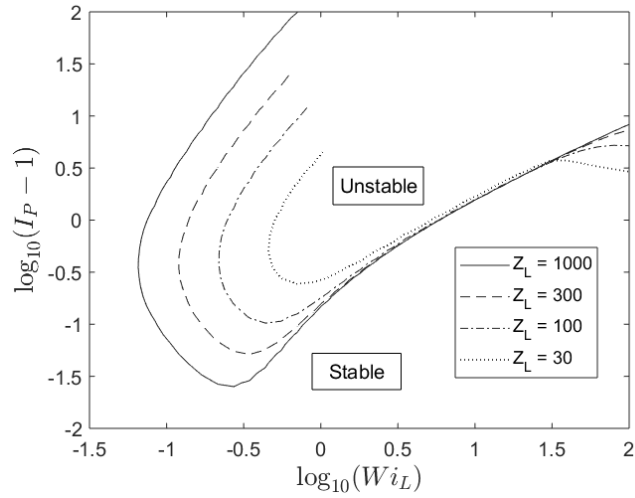

Figure 1: Neutral stability boundaries to SID for bi-disperse blends over a range of compositions (all with $\phi_L = 0.5$) and shear rates. At very low polydispersity, the melt is stable at all shear rates. However, above a polydispersity index of $I_P - 1 \sim \mathcal{O}(1/Z_L)$ a range of shear rates appears for which the flow is linearly unstable to SID. The lower end of this unstable window decreases with increasing polydispersity, but the upper end is roughly fixed over a very wide range of $Z_L$.

Figure 1 shows that at very low polydispersity, there is insufficient contrast between the two components to driven any segregation behavior – after all, a melt of identical components cannot exhibit a demixing instability. However, at higher polydispersity $I_P - 1 \sim \mathcal{O}(1/Z_L)$ a range of shear rates appears for which the flow is linearly unstable to SID. This lower bound polydispersity for SID coincides with the point where elastic forces ($k_B T$ per entanglement) scaled by the contrast between chains ($I_P - 1$) exceed the entropic forces for mixing ($k_B T$ per chain). When SID first occurs at low polydispersity, the unstable window is centered about $Wi_L \sim 0.3$ – lower shear rates are stable because elastic stresses are too weak to drive migration and higher shear rates are stable because elastic forces create enhanced mixing.

A region of flow-enhanced mixing is visible in Figure 1 insofar as the high $Wi_L$ boundary of the unstable region is essentially the same for all values of $Z_L$. Over the range of $Z_L$ shown in Figure 1, the stability boundaries all coincide at the $Wi_L$ where the effect of flow switches from destabilizing to stabilizing.

The enhanced mixing phenomena at $Wi_L > 1$ can be explained roughly as follows: for $Wi_L > 1$, well entangled polymers shear thin due to increasing alignment in the flow direction. The long chains are always more aligned with the flow than short chains, and therefore their stress is less sensitive to the changes in shear rate that occur when the melt composition is perturbed. The short chains, which are less aligned with the flow, experience a comparatively large change in stress when the composition is perturbed. Since the short chains are more stressed where they are most concentrated (where the shear rate has increased) elastic forces promote a migration of short chains towards regions of higher molecular weight, thereby enhancing mixing.

In two-fluid models of polymer *solutions*, it has been found that shear thinning at $Wi > 1$ does not favor mixing and is instead crucial to the mechanism of SID [14]. The difference is partly explained by the fact that in a polymer solution, unlike a low polydispersity blend, the low viscosity component (solvent) has no appreciable contribution to the stress. Indeed, Figure 1 shows that when the short chain viscosity is about twenty times smaller than the long chain viscosity ($I_P \sim 1.3$) the unstable window extends to cover the shear thinning regime, $Wi_L > 1$, as the short chains no longer carry a substantial portion of the stress. In this range, decreasing $Z_S$ primarily serves to increase the entropic forces favoring mixing, and so for $I_P > 2$ we find that increasing polydispersity makes the system more stable at low $Wi_L$.

In Figure 2, we repeat the analysis of Figure 1 using the Doi and Onuki description for $\mathcal{F}_{DO}$ as defined in equation (4). For the range of $Wi_L < 10$, there appears to be good qualitative agreement between the two frameworks. Both frameworks predict that SID is absent at sufficiently low polydispersities, and with increasing polydispersity SID first emerges at shear rates below $Wi_L < 1$. At higher polydispersities, the unstable window for SID shifts into the shear thinning regime, $Wi_L > 1$. At higher $Wi_L$ and higher polydispersity the predictions of these two frameworks begin to diverge. The qualitative departure at high $Wi_L$ is similar to what has been found for polymer solution models of SID with the same conflict in implicit/explicit treatments of an elastic free energy [14] [27]. Because we cannot yet a-priori determine which of the two frameworks is best, the remainder of this report will be focused on predictions in parameter spaces where there is good agreement. Unless otherwise stated, all predictions to follow will employ an explicit treatment of the elastic free energy, per equation (7).

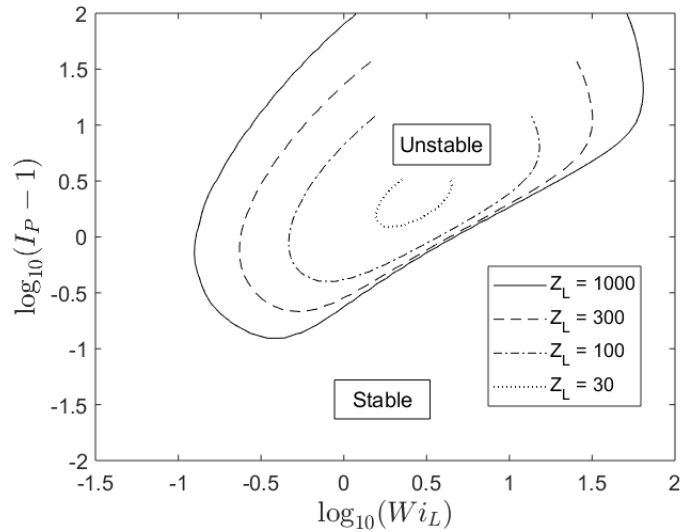

*Figure 2: Neutral stability boundaries to SID for bi-disperse blends over a range of compositions (all with $\phi_L = 0.5$) and shear rates, now using the Doi and Onuki description for the elastic forces driving polymer migration. At very low polydispersity, the melt is stable at all shear rates. However, above a polydispersity index of $I_P - 1 \sim \mathcal{O}(1/\sqrt{Z_L})$ a range of shear rates appears for which the flow is linearly unstable to SID. The unstable window appears to show signs of closing at high polydispersity, due to the fact that short chains provide a stronger entropic incentive for mixing.*

In the range of $Wi_L < 10$, two subtle but important differences between Figure 1 and Figure 2 are (1) the range of unstable flow conditions is slightly smaller in the Doi-Onuki approach and (2) the lowest polydispersity for SID in Figure 2 scales as $I_P - 1 \sim \mathcal{O}(1/\sqrt{\bar{Z}})$. The increased stability of homogeneous flow in the Doi/Onuki framework is partly explained by the fact that the leading order terms for stress-induced migration vanish at $\mathcal{O}(I_P - 1)$ for $I_P - 1 \ll 1$.

The preceding calculations present a relatively complete picture for how SID is influenced by flow conditions and melt composition in blends with $\phi_L = 0.5$. In Figure 3, we fix $Z_L = 100$ and extend this analysis to consider blends of differing long chain volume fraction, $\phi_L = 0.2, 0.4, 0.6, 0.8$. In the range of $\phi_L$ considered here, we find that increasing the long chain fraction mostly serves to shift the unstable window to cover a lower range of polydispersity. At even higher values of $\phi_L$, the unstable window will eventually close off altogether – there are so few short chains that in order to produce any meaningful rheological contrast, the short chains must be so short that entropic mixing forces dominate over stress-induced migration effects.

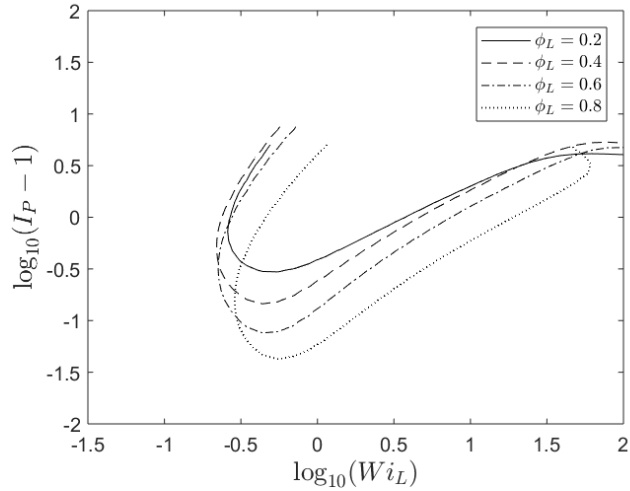

Figure 3: Neutral stability boundaries to SID for bi-disperse blends with $Z_L = 100$ and varying $\phi_L$ over a range of compositions and shear rates. In the range of $\phi_L$ considered here, we see that increasing $\phi_L$ primarily serves to shift the unstable window to lower polydispersity and (to a lesser extent) higher shear rate. At even higher values of $\phi_L$, increasing $\phi_L$ will close off the unstable window altogether.

To this point, we have not yet considered any chemical contrast between chains, and all of the above figures consider only the case of $\chi_{LS} = 0$. For $\chi_{LS} \neq 0$, it is our view that the results are generally less interesting: whereas the elastic forces driving chain migration are of order $k_B T$ per entanglement, the enthalpic forces favoring mixing are of order $\chi_{LS} k_B T$ per monomer. If $\chi_{LS} N_e \sim \mathcal{O}(1)$, enthalpic mixing terms will dominate over elastic migration terms at all shear rates (i.e. shear flow by itself is probably not a practical tool to compatibilize immiscible blends). The more interesting physics of flow-induced mixing/demixing would only be visible for $|\chi_{LS}| N_e < 1$, at which point we would trivially see that a finite $\chi_{LS}$ will promote mixing/demixing at all shear rates to contract/broaden the neutral stability boundaries as shown in Figure 1 and Figure 3.

In support of the discussion in the preceding paragraph, we present calculations for a bidisperse blend with $Z_L = 100$ and $\phi_L = 0.50$ and varying $\chi_{LS} N_e = -0.05, 0, 0.02, 0.05$. We also remind the reader that so far we have not accounted for how a contrast in segment length or entanglement molecular weight effects the thermodynamics of mixing, the elastic free energy, the coupling friction, or the stress relaxation dynamics – those details are much more complex and must be deferred to future studies. In Figure 4 below, we see that small increases/decreases from $\chi_{LS} = 0$ simply broaden/contract the unstable region observed for $\chi_{LS} = 0$. Around $\chi_{LS} N_e \sim 2/\bar{Z}$, exemplified by the $\chi_{LS} N_e = 0.05$ case, the stability boundary breaks into two separate curves – indicating that low polydispersity systems are no longer thermodynamically stable at rest.

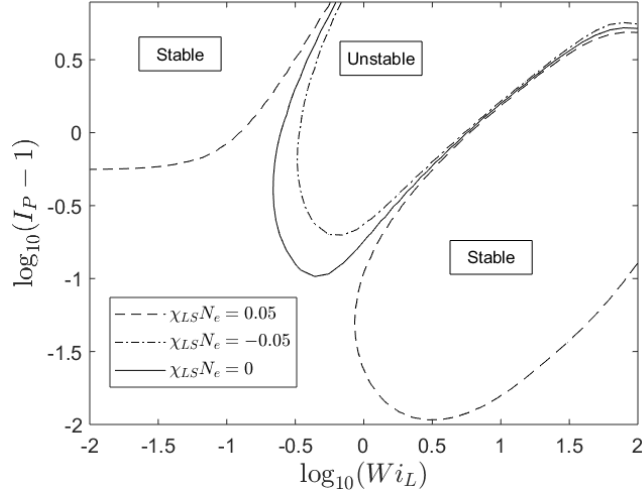

*Figure 4: Neutral stability boundaries to SID for bi-disperse blends with $Z_L = 100$ and $\phi_L = 0.5$ over a range of chemical contrasts between components, $\chi_{LS} = -0.05, 0, 0.02, 0.05$. In the range of $\chi_{LS}$ considered here, we see that increasing $\chi_{LS}$ simply broadens the unstable window observed at $\chi_{LS} = 0$. When $\chi_{LS}N_e \sim \mathcal{O}(1)$, flow cannot overcome the system's thermodynamic preference for mixing/demixing.*

While it is not shown in Figure 4, we have found that for $\chi_{LS}N_e = 1$, homogeneous flow is unstable across the entire range of flow rates and compositions shown in Figure 4. This is not an unexpected result; the elastic forces favoring mixing are at most $k_BT$ per entanglement while the enthalpic forces favoring demixing are $k_BT\chi_{LS}$ per monomer. When $\chi_{LS}N_e = 1$, the enthalpic driving force for segregation is simply too large to be overcome by flow-induced stresses.

Finally, we note that the existence of linearly stable flow solutions for $\chi_{LS}N_e = 0.05$ and $Wi_L > 1$ is a necessary but not sufficient condition for flow-induced mixing when the initial condition is a fully segregated state. It is possible, for example, that a demixed flow solution exists and is locally stable to small perturbations under flow conditions with the same overall $Wi_L$. Further studies would be needed to establish predictions for flow-induced mixing, but this lies beyond the scope of our present work.

## Section IV.B: Linear Stability: Finite wavelengths

In the preceding section, we considered long wavelength perturbations as a means of probing a macroscopic system's absolute stability to SID. In this section, we will consider a system that falls within the unstable window of Figure 1 and describe its response to perturbations of finite wavelengths. A principal result for this section is that when the system is unstable to SID, the fastest growing linear mode for SID has a microscopic wavelength comparable to the typical polymer coil size.

The wavelength dependence of a growth rate for SID has been previously discussed in studies of wormlike micelles [25] and polymer solutions [14] [27]. When a SID instability exists, the growth of a perturbation is limited by mass transport at long wavelengths and suppressed at short wavelengths by the 'interfacial' osmotic term. For intermediate wavelengths, there is a 'fastest growing linear mode' that balances these limitations. When the growth rate of the fastest growing linear mode approaches the typical stress relaxation time, one often finds that demixing (which deforms the fluid) cannot proceed faster than the rate at which the 'migration induced stresses are able to relax. This leads to a 'plateau' in the growth rate vs wave-number centered about the fastest growing mode [25] [14]. We observe the same overall behavior for SID in our polymer blend model.

To confirm that our blend model qualitatively reproduces these features known to SID in polymer solutions, we discuss the results for a blend of $Z_L = 100$, $Z_S = 20$, and $\phi_L = 0.5$ ($\bar{Z} = 60, I_P = 1.8$) at $Wi_L = 1$. These results are representative of what is predicted for most other blend compositions whenever the system is unstable to SID. Figure 5 shows the growth rate, $\sigma$, as a function of the wavenumber $k$ over the range $k\bar{R}_g = 10^{-2} - 10^2$. Recall that the growth rate $\sigma$ has units of $1/\tau_D(\bar{Z})$ and the wavenumber $k$ has units of $1/H$ in our non-dimensionalization scheme.

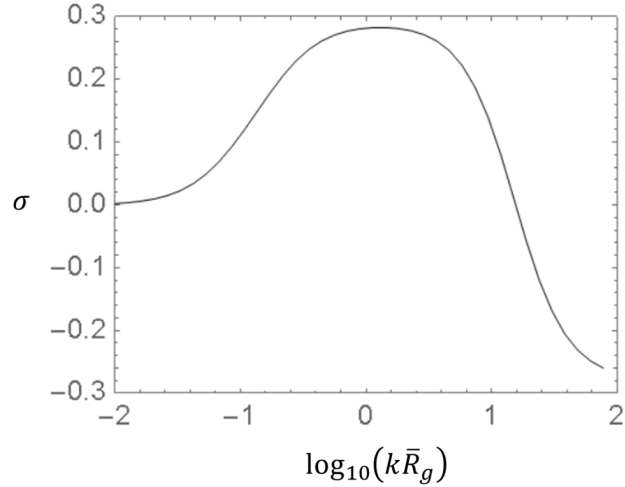

Figure 5: Growth rate $\sigma$ (in units of $1/\tau_D(\bar{Z})$) for a blend of $Z_L = 100$, $Z_S = 20$, and $\phi_L = 0.5$. At low wavenumbers, $k\bar{R}_g \ll 1$, the perturbation is unstable but its growth is diffusion limited, $\sigma \sim (k\bar{R}_g)^2$. At much higher wavenumbers, $k\bar{R}_g > \bar{Z}^{1/2}$, the perturbation's growth is suppressed due to capillary forces. In between, there is a range of wavenumbers for which the growth rate roughly plateaus, with a maximum growth rate around $k\bar{R}_g \sim 1$.

At low wavenumbers, $k\bar{R}_g \ll 1$, the perturbation is unstable, $\sigma > 0$, but its growth is diffusion-limited, $\sigma \sim (k\bar{R}_g)^2$. At higher wavenumbers, $k\bar{R}_g > \bar{Z}^{1/2}$, the perturbation wavelength becomes comparable to the tube diameter, and it decays due to interfacial forces manifested by the square gradient term in the mixing free energy (3) (note the sign change in $\sigma$ around $k\bar{R}_g \sim 10$). For intermediate wavenumbers there is a rough 'plateau' in the growth rate with a 'most unstable mode' appearing around $k\bar{R}_g \sim 1$ and $\sigma \sim 0.3$. The 'plateau' in growth rate appears because demixing is limited by the stress relaxation time, which has no wavenumber dependence [27]. For shear rates closer to the neutral stability boundary $\sigma_0 = 0$, the most unstable mode shifts to smaller wavenumbers and slower growth rates.

It is worth noting that, except for very near the neutral stability boundary $\sigma_0 = 0$, the fastest growing linear mode for SID is typically found at wavelengths comparable to the polymer coil size, $k\bar{R}_g \sim \mathcal{O}(1)$. Thus, it is extremely unlikely that the earliest measurable signature of SID in a polymer blend would bear resemblance to a shear banding instability with two distinct bands spanning a macroscopic gap.

### Section IV.C: Nonlinear Dynamics of SID

To conclude our discussion of SID in bi-disperse polymer blends, we will consider predictions for the non-linear evolution of a SID instability.

The non-linear dynamics of SID in a viscoelastic two-fluid model have been previously described in detail for wormlike micelles, polymer solutions, and blend-like systems [27] [10] [16] [11]. A survey of these works reveals a number of repeated themes, which we will summarize briefly. Linear stability analysis is insightful for describing the initial stages of demixing [25], but at later times the amplitude of the perturbation saturates and a stratified or 'banded' structure develops [28] [14] [16]. At still later times, calculations of SID using the '1D' approximation show that the banded structure coarsens via a process resembling Oswald ripening [20]. Other coarsening mechanisms have been observed in higher dimensions [11], but the overall structure remains stratified in the flow gradient direction. Any change to material parameters that would favor demixing in the linear stability analysis (e.g. weaker mixing forces relative to elastic forces) also tends to favor increased segregation in the final banded state.

Our study of SID in polymer blends reproduces all of the above observations. In this section, we will discuss results for the non-linear dynamics of SID in a blend of $Z_L = 100$, $Z_S = 30$, and $\phi_L = 0.5$ ($\bar{Z} = 65$, $I_P = 1.4$). This choice of parameters represents, for example, a 50/50 blend of polybutadiene, with the long and short chain species having molecular weights of 180kg/mol and 54kg/mol, respectively. For the imposed deformation, we have chosen to employ a constant shear stress ($\sigma_{xy} = 0.16$ in units of the shear modulus $G$) instead of a constant shear rate, but this does not qualitatively change any feature of the resulting calculations [20][4].

The results we describe in this section are qualitatively representative of what is predicted for other blend compositions and shear rates where SID occurs. Results are also given to show how changes in blend composition and applied stress quantitatively influence the final banded state.

To reduce the impact of boundary effects in our simulations, we choose a small value of $\bar{R}_g = 0.10$ so that the typical polymer coil is roughly 10 times smaller than the gap dimension. We also employ Lees Edwards boundary conditions [10], which can be understood as an analogue to periodic boundary conditions for systems under steady shear (e.g. the velocity is not periodic, but the shear rate is). There are two main advantages to Lees Edwards boundary conditions in these simulations: first, Lees Edwards boundary conditions allow us to study a portion of the 'bulk' flow domain without influence from solid surface boundary conditions. Second, Lees Edwards boundary conditions permit the use of fast and accurate pseudospectral Fourier methods in our numerical calculations.

Our simulations begin from an initial condition in which the polymer configuration is relaxed, but the concentration profile is perturbed from its homogeneous state by random Gaussian noise of amplitude $10^{-6}$. As we describe how the system evolves in time, we will focus on the changes that occur in the molecular weight distribution. At steady state, we will also discuss changes in the velocity gradient.

At short times ($t = 8.9$ in the figure below), we find that the system selects a preferred wavelength for demixing from the initial random noise. This is also observed for shear induced demixing in polymer solutions, and can be anticipated from Figure 5 and the surrounding discussion. The preferred wavelength for demixing observed here is roughly half of the system size, and so we conclude that finite size effects are not responsible for the selection of a preferred wavelength during demixing.

---

[4] We do note, however, that non-dimensionalizing for stress-driven flows requires a new choice of characteristic velocity scale ($u_C = H/\tau_C$) since the actual flow velocity is no longer known. This choice of $u_C$ ensures that the dimensionless velocity difference across the simulated domain is the Weissenberg number for the flow.

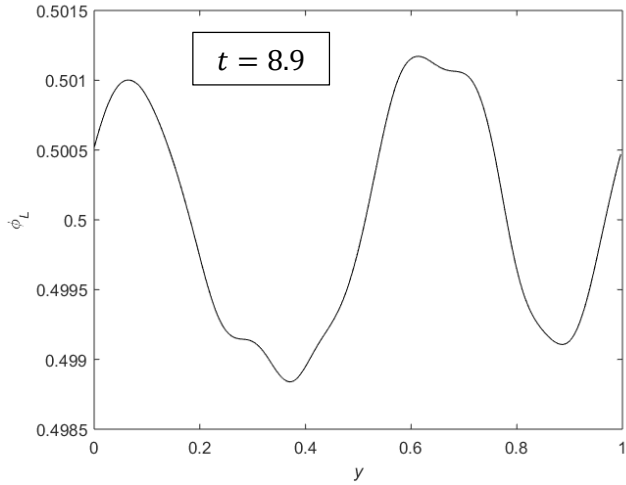

*Figure 6: Beginning from an initial concentration profile of small amplitude random noise, a blend of $\bar{Z} = 65, I_P = 1.4$, and $\phi_L = 0.5$ will spontaneously begin demixing when deformed by a constant dimensionless shear stress, $\sigma_{xy} = 0.16$. The amplitude of the initial random noise ($10^{-6}$) is much smaller than the amplitude to which the perturbation has grown by time $t = 8.9$ ($10^{-3}$), and the instability has selected a fastest-growing mode. Note that the fastest growing mode has a wavelength that is smaller than the system size (finite size effects are not responsible for the selection of a finite wavelength mode) and is slightly longer than the typical polymer coil size (here, $\bar{R}_g = 0.10$).*

At longer times ($t = 20$ In the figure below) we find that some of the 'ridges' in the concentration profile have smoothed out, and the concentration profile mostly alternates between nearly homogeneous regions that are at high/low $\phi_L$. The width of the interface $\ell$ between the two nearly homogeneous regions is set by the length at which interfacial forces, $k_B T/b\ell^2$, matches the driving force for segregation, which is at most $k_B T/b^3 N_e$ for homologous chains. Thus, the width of the interface can be no smaller than the tube diameter $\ell > bN_e^{1/2}$.

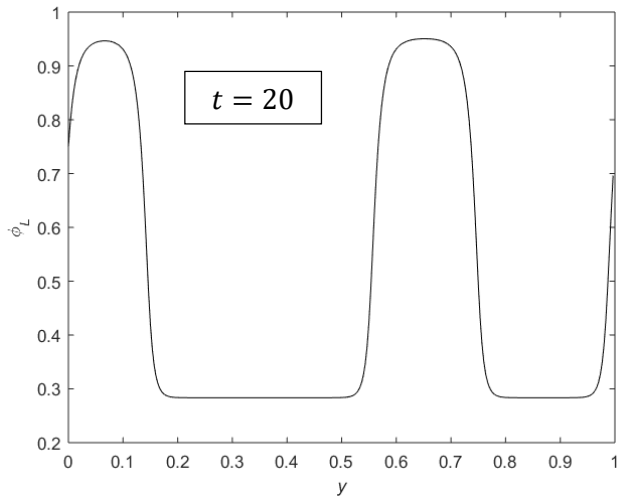

*Figure 7: The system from Figure 6 continues to demix until non-linear terms cause the amplitude of the perturbation to saturate. Here, we find that by $t = 20$ the system has segregated into bands that are rich in long chains ($\phi_L = 0.95$) and bands that are poor in long chains ($\phi_L = 0.3$). Recall that the average composition of the system is $\phi_L = 0.5$. Finally, we note that the interface between bands (and the bands themselves) have a width several times larger than the typical coil size, $R_g$.*

At longer times, the system undergoes a kind of Oswald ripening to reduce its interfacial free energy. The steady state concentration profile, shown below, contains two distinct bands of high/low concentration connected by relatively sharp interfaces. This ripening process is very slow, but ultimately it is only limited by the size of the system and the chosen boundary conditions. Below, we show the steady state concentration profile. Recall that the concentration has periodic (Lees Edwards) boundary conditions:

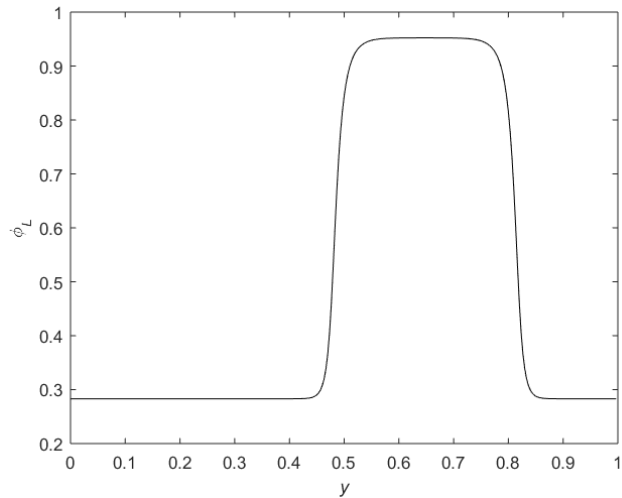

*Figure 8: The system from Figure 7 coarsens towards macroscopic demixing at steady state, at least to the extent permitted by the finite system size and periodic boundary conditions.*

Unsurprisingly, the regions at high/low $\phi_L$ are also at low/high shear rates relative to the mean. Technically, this amounts to shear banding, but the shear bands predicted here have a width comparable to the polymer coil size and would be too small for conventional PTV imaging methods.

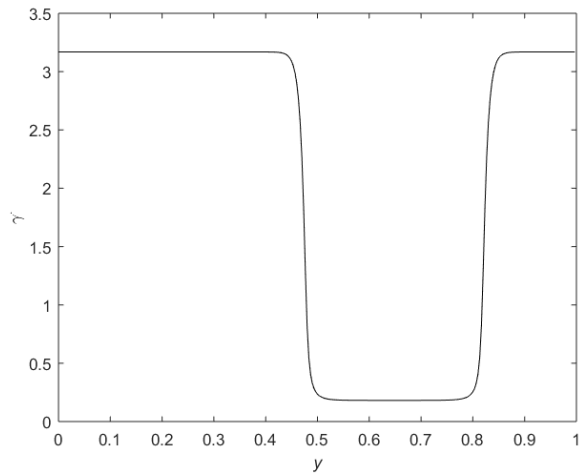

*Figure 9: The system from Figure 8 possesses an inhomogeneous velocity profile to complement the inhomogeneous concentration profile. Regions with fewer long chains (being less viscous) will have a higher shear rate, and regions with more long chains (being more viscous) will have a lower shear rate. For constant shear stress boundary conditions, the shear rate $\dot{\gamma}$ is defined as $\dot{\gamma} = \partial u_T / \partial y$, and the average shear rate defines an effective Weissenberg number ( $Wi = 2.1$ in this case).*

At steady state, the differences in composition and shear rate can be very significant. In this case, the low shear band is ~95% long chains (recall that the overall blend composition is 50/50) and the shear rate in the two bands differs by more than an order of magnitude. This can be contrasted with prior findings for polymer solutions, in which the composition differences between the bands typically varied by less than 10% [14] [27]. At low polydispersity and high entanglement numbers, the rheological contrast needed for SID may be less than the contrast needed to distinguish the rheology of the two components in isolation. When this occurs, SID may occur but go unnoticed for rheological purposes.

The non-linear dynamics of SID at other blend compositions and shear rates follow qualitatively similar trajectories. Taking the preceding simulation as a reference case, we consider how changes in $I_P$ and $\bar{Z}$ effect polymer segregation in the final steady state. In each case, we maintain fixed $\phi_L = 0.5$ and $\sigma_{xy} = 0.16$. First, we maintain $\bar{Z} = 65$ and decrease the polydispersity index to $I_P = 1.2$. This decreases the rheological contrast between long and short chains, which reduces the strength of the elastic forces promoting demixing. As a result, the differences in composition (and shear rate) become less pronounced between the two bands. Second, we maintain $I_P = 1.4$ and increase $\bar{Z}$ to $\bar{Z} = 100$. The rheological contrast between chains is effectively unchanged, but now the chains are all much larger and their mixing free energy is reduced relative to the elastic free energy. At steady state, differences in composition become even more pronounced before thermodynamic mixing forces can balance the elastic forces driving segregation.

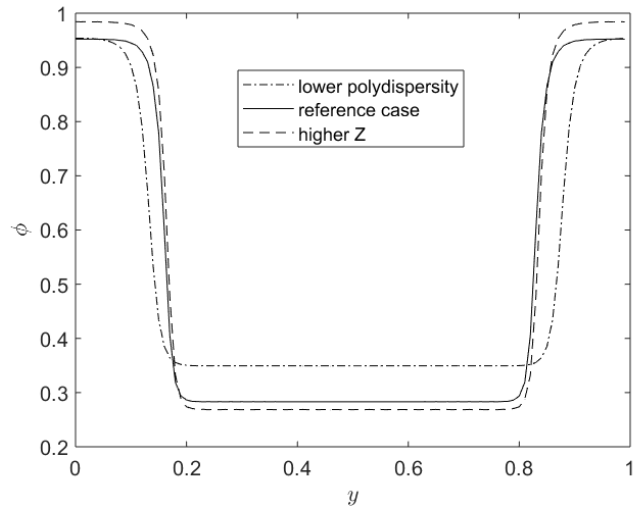

*Figure 10: Composition profiles for inhomogeneous steady states with slightly different parameters and the same imposed shear stress, $\sigma_{xy} = 0.16$. Relative to the reference case (see Figure 8) we see that increasing the entanglement number $\bar{Z}$ (dashed line) leads to increased segregation of long and short chains in the demixed steady state. Likewise, decreasing the polydispersity index (dash-dotted line) leads to decreased segregation of long and short chains.*

Finally, we also consider the effects of changing the applied shear stress, $\sigma_{xy}$. Since increasing the applied shear stress increases the overall shear rate (even when the system has demixed), we expect that higher shear stresses will have a larger proportion of the high shear band spanning the system. In Figure 11, we return to our reference blend composition of $I_P = 1.4, \bar{Z} = 65$, and $\phi_L = 0.5$. Here, we show that decreasing the shear stress from $\sigma_{xy} = 0.16$ to $\sigma_{xy} = 0.11$ decreases the fraction of high shear band spanning the system. We also show that changes in the applied shear stress lead to changes in the shear rates within the 'bands', which is different from the usual 'lever rule' behavior reported in most experimental observations of macroscopic shear banding.

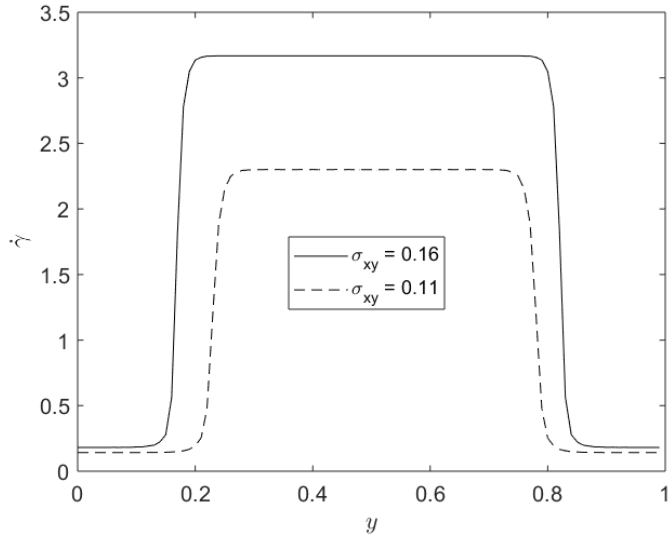

*Figure 11: Shear rate profiles for inhomogeneous steady states with slightly different imposed shear stresses ($\sigma_{xy} = 0.16$ for the solid curve and $\sigma_{xy} = 0.11$ for the dashed curve) but the same material parameters ($\bar{Z} = 100$, $I_P = 1.4$, and $\phi_L = 0.5$). We see that increasing the applied shear stress leads to an increase in the shear rate within each 'band', and also to an increase in the fraction of space occupied by the higher shear band.*

Overall, the results of our non-linear simulations are qualitatively consistent with what has previously been observed for the non-linear dynamics of SID instabilities [14] [28] [29] [16] [20]. We also remind the reader that the fastest growing linear mode of a SID instability is typically microscopic in scale, $k\bar{R}_g \sim \mathcal{O}(1)$, and our simulations confirm that coarsening to larger lengths is a slow process. As such, SID is probably not responsible for any observations of macroscopic shear banding in commercially (i.e. rapidly) processed polymer melts (e.g. [30]). Instead, observing the effects of SID in chemically similar chains will likely require something like a rheoSANS study of deuterated bi-disperse polymer blends in a recirculating shear cell.

## Section V: Calculations for polydisperse blends

In section IV, we showed that SID can occur in bidisperse blends of chemically identical polymers simply due to a constrast in the rheology of the two components. Polydisperse blends present more of a conceptual challenge – if the molecular weight distribution is smeared from two delta functions (bi-disperse) into something smooth and uni-modal (e.g. log-normal), do chains of intermediate size suppress demixing instabilities by screening the rheological contrast of longer and shorter chains?

According to the predictions of our model, the answer is no; predictions of shear induced demixing are not a peculiarity of bi-disperse blends, rather they are a feature of polydisperse blends in general.

## Section V.A: Conditions for SID

For the present study of SID in polydisperse blends, we limit our inquiry to blends of log-normal molecular weight distributions:

$$\phi(z) = [2\pi z^2 \ln(I_P)]^{-1/2} \exp\left[-\frac{\ln(z\sqrt{I_P})^2}{2\ln(I_P)}\right]$$

The log-normal molecular weight distribution has no basis in first-principles reaction polymerization kinetics, but it is a useful approximant to the molecular weight distributions seen in industrial practice [31]. In *Figure 12*, we show the neutral stability boundary $\sigma_0 = 0$ over the range $Wi = 0.02 - 5$ for log-normal polydisperse blends with mean entanglement numbers $\bar{Z} = 30, 100, 300, 1000$ and polydispersity index ranging from $I_P - 1 = 0.01 - 0.4$. SID is not observed for the $\bar{Z} = 30$ blends for this range of polydispersity and shear rate. Overall, for the range of polydispersity and shear rates considered here, these neutral stability boundaries are qualitatively similar to the bi-disperse results of Figure 1. Similar results have also been obtained, for example, with a Flory distribution for the molecular weight [26].

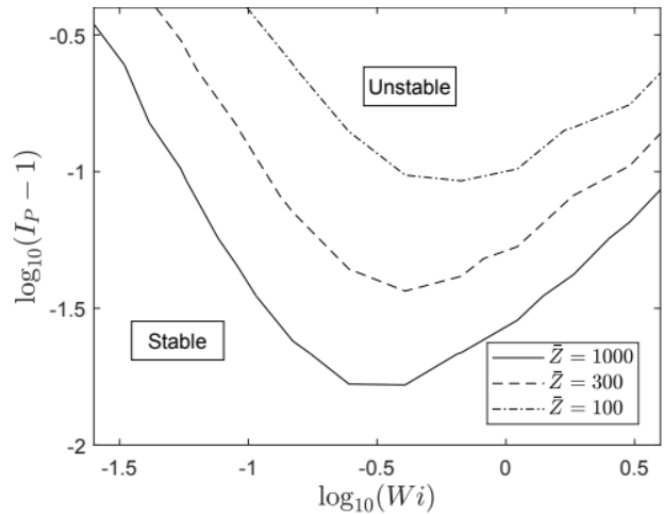

*Figure 12: Neutral stability plot for SID in polydisperse blends of log-normal molecular weight distribution. In qualitative agreement with Figure 1, we see that blends of higher entanglement number show SID at lower polydispersity.*

At higher polydispersity, the lower critical shear rate for SID continues to decrease with increasing polydispersity. This is qualitatively different from what was seen for bi-disperse blends because in a polydisperse blend it is possible for chains in the low molecular weight tail (which entropically favor mixing) to remain well mixed while fractionation occurs within the high molecular weight tail.

## Section V.B: Predicted fractionation effects

If the homogeneous steady flow solution has a molecular weight distribution $\phi(t, y, z) = \phi_0(y, z)$ then we one can apply an infinitesimal perturbation, $\phi(t, y, z) = \phi_0(y, z) + \delta\phi_0(z)\exp(iky + \sigma t)$, to study the linear stability to demixing. Here, we have simplified the form of the imposed perturbation to only consider plane wave perturbations to the composition that grow or decay exponentially in time. When SID occurs in a polydisperse melt, the resulting fractionation of the molecular weight distribution is potentially non-trivial, and the form of $\delta\phi_0(z)$ is not immediately obvious.

By evaluating the eigenvalues of an unstable mode in the $k\bar{R}_g \ll 1$ limit, we can gain a sense for the fractionation that emerges upon demixing. At very low polydispersities, $I_P - 1 \ll 1$, this analysis reveals a simple trend: chains that are longer/shorter than the average size migrate opposite one another, and the speed of migration for a given chain is proportional to the difference in molecular weight between that chain and the 'average' chain (i.e. $\delta\phi_0(z) \sim \phi(z)(z-1)$ for any $Wi$ where SID occurs) [26]. At higher polydispersity, however, the fractionation induced by SID is far more complex.

As a representative example, we present a set of results for the fastest-growing perturbation to the molecular weight distribution $\delta\phi_0(z)$ in the limit $k\bar{R}_g \ll 1$, where $z = Z/\bar{Z}$. For this example case, we consider a blend of $\bar{Z} = 100$, $I_P = 1.5$, and compare results at varying shear rates, $Wi = 0.3 - 5.5$. Figure 13 shows that that SID can lead to very complex changes in the molecular weight distribution – some short chains move towards region of higher average molecular weight, and some move opposite (and likewise for the long chains). For the range of $Wi$ considered here, increasing the shear rate seems to 'shift' the perturbation in the direction of decreasing molecular weight. At this time, we cannot provide a complete explanation for the complexities of these predictions, but we believe it is a subject that certainly warrants further inquiry.

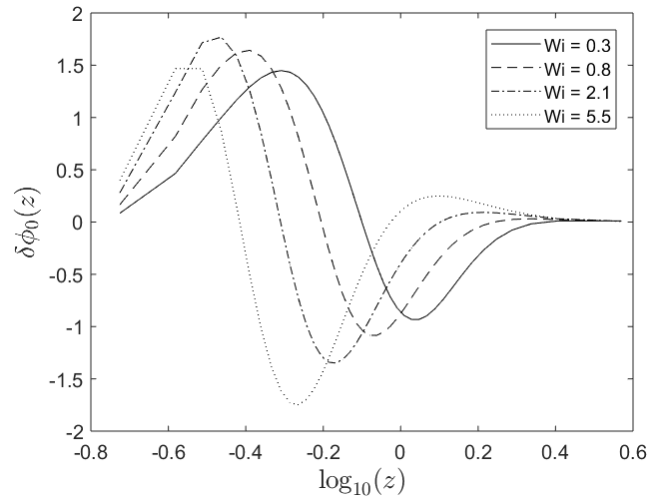

Figure 13: For a log-normal blend of $\bar{Z} = 100$ and $I_P - 1 = 0.5$ ($N = 40$) we compare predictions (at varying $Wi$) for the fastest-growing linear perturbation to the molecular weight distribution in the limit $k\bar{R}_g \ll 1$. The perturbation to the molecular weight distribution is complex and varies with $Wi$.

Finally, whereas bi-disperse blends were able to show flow-induced stabilization against demixing for a range of $Wi > 1$ and $I_P < 2$ (c.f. Figure 4), the same predictions are less likely to be observed in mixtures of polydisperse blends. For example, when each polydisperse blend is independently susceptible to demixing, it seems unlikely that the combined mixture could somehow remain compositionally homogeneous.  As such, shear flow alone may be ineffective for compatibilizing immiscible blends with polydisperse components (even for very small chemical contrasts, $\chi N_e \ll 1$).

## Section VI:  Summary and Conclusions

In light of recent developments in the constitutive modelling for bi-disperse and polydisperse entangled linear polymers [19], we have developed an improved 'multi-fluid' generalization of the classic two-fluid approximation for studying inhomogeneous polymers in flow.  As a preliminary application, we considered shear induced demixing phenomena in chemically homologous linear polymer blends with bi-disperse and log-normal molecular weight distributions.

After non-dimensionalizing the governing equations, we found that the system's response to shear flow is mainly governed by just three dimensionless parameters:  (1) $\bar{Z}$, the mean entanglement number of the blend (2) $I_P$, the polydispersity index of the blend, and (3) $Wi$ the Weissenberg number of the flow. Other parameters (namely $\bar{R}_g$ and $\eta$) play a much smaller role by comparison.

Considering the stability to SID for bi-disperse blends, we found that the lower bound polydispersity for SID coincides with the point where elastic forces ($k_BT$ per entanglement) scaled by the contrast between chains ($I_P - 1$) exceed the entropic forces for mixing ($k_BT$ per chain).  At low polydispersity, SID first appears at relatively low shear rates, $Wi \sim 0.3$, when elastic stresses are moderate but not yet shear thinning.  With increasing polydispersity, the unstable range initially broadens to span higher/lower $Wi$ – however, eventually this trend reverses on the lower range of $Wi$ once the entropy of mixing for short chains becomes more significant.

With regard to the dynamics of SID, we observe results that are qualitatively consistent with what has been previously shown for SID in entangled polymer solutions.  The system demixes into a microscopically 'shear banded' structure with a band thickness comparable to the typical polymer coil size.  Over much longer time-scales the bands merge together, with the time needed to attain a macroscopically banded steady state likely exceeding cycle times in commercial polymer melt processing.

With regard to the composition and contrast exhibited by a fully demixed flow, we find that the contrast between the layers depends on the contrast in the component polymers – in other words, while it may be theoretically possible for a blend of $\bar{Z} = 100$ and $I_P = 1.1$ to demix, the polydispersity index is so low that demixing would be unlikely to produce a notable change in the observed rheology. On the other hand, when the polydispersity index is not small, the contrast between bands can be substantial – for $\bar{Z} = 100$ and $I_P = 1.4$, our simulations suggest that the shear rate in the two bands can differ by nearly an order of magnitude!

Finally, we confirmed that many of the results for bi-disperse blends are transferrable to a study of polydisperse blends with log-normal molecular weight distributions of low polydispersity index.  A principal distinction between the two molecular weight distributions is that the continuous molecular weight distribution allows an infinite set of possible perturbations to the molecular weight distribution while the bi-disperse blend allows only one.  As such, it seems that the polydisperse blend will show more susceptibility to demixing at high polydispersity (the short chains can remain homogeneous while long chains demix amongst themselves) and less of a tendency for enhanced mixing (there will always be some slow-decaying perturbation even when no perturbation leads to demixing).

A significant limitation of the present work is that although we have performed calculations across a wide range of bi-disperse blend compositions, the validity of the SRDP constitutive model is only well-established for blends in region 3 of the Viovy diagram [32] [19]. In a forthcoming publication, we will discuss appropriate two-fluid models for the adjoining quadrants (regions 2 and 4) and compare model predictions at the region boundaries. This broader study affirms the main points of the work presented here.

Flow induced fractionation in melts of polydisperse entangled polymers is a rich and interesting subject that extends well beyond the topic of SID. We hope that the study presented here will motivate additional interest in future experimental and theoretical studies.

# Appendicies

## Section A.1: Elastic free energy derivation

At the request of Daniel Read these sections do not appear in the Arxiv version of our report. His derivation of an elastic free energy in entangled polymer melts will be the main subject of a forthcoming publication.

## Section A.2: Dimensionless equations (simple shear)

In this section, we present the full set of dimensionless governing equations for the SRDP multi-fluid model in 1D simple shear flow. The geometry is defined by a pair of infinite parallel plates separated by a distance $H$ in the $y$-direction, and an $N$-component SRDP blend fills the gap in between. The top plates is translated in the $x$-direction at a velocity $U$ relative to the bottom plate, and the flow is assume to be uniform in the direction of flow ($x$-direction) and vorticity ($z$-direction).

Proceeding toward the final dimensionless governing equations, we denote scalar components of tensors by their subscripts (e.g. the $xy$ component of the tensor $\boldsymbol{Q}_{ij}$ is given by $Q_{ij,xy}$), and species velocities in the flow and flow gradient direction are given by $u_m$ and $v_m$, respectively.

The characteristic scale for elastic stresses is the shear modulus $\sigma_C \sim G$ ($k_B T$ per entanglement) while the characteristic scale for the chemical potential $\mu = \delta F/\delta \phi$ is $\mu_C \sim G/\bar{Z}$ ($k_B T$ per chain), where $\bar{Z} = \sum_m \phi_m Z_m$ is the average number of entanglements per chain in the homogeneous base state. The characteristic length-scale in the $y$-direction is the gap size $H$. The characteristic scale for velocities in the flow direction is given by the velocity difference between the upper and lower plates, $u_C \sim U$. The characteristic time-scale is the reptation time for a chain of average length, $\tau_C = \tau_D(\bar{Z})$. The characteristic velocity in the flow gradient is set so that the characteristic scale of drag forces matches the characteristic scale of chemical potential gradients, $v_C \bar{\zeta} \sim \mu_C/H$, where $\bar{\zeta} = \sum_i \phi_i \zeta_i$ is the average drag coefficient in the homogeneous base state.

Given the definitions of characteristic scales and dimensionless parameters above, we can now state the dimensionless governing equations for multi-fluid SRDP fluids in 1D simple shear flow. From this point forward, we explicitly evaluate all functional derivatives in terms of the previously defined Free energy expressions, and all equations are assumed to be dimensionless.

The dimensionless equations for our multi-fluid model of polymer blends in 1D shear flow are as follows:

The continuity equation for the volume fraction $\phi_m$ of chain $m$ is given by:

$$\frac{\partial \phi_m}{\partial t} = -\bar{R}_g^2 \frac{\partial}{\partial y}(\phi_m v_m) \qquad (18)$$

where $\bar{R}_g = a\bar{Z}^{1/2}/H$ is the typical polymer coil size relative to the gap dimension, and $\bar{Z} = \sum_i \phi_i Z_i$ is the mean entanglement number for all chains in the system. In a 1D simple shear flow, the incompressibility equation is equivalent to a uniformly zero volume averaged velocity in the flow gradient direction:

$$\sum_i \phi_i v_i = 0 \qquad (19)$$

It is worth pointing out that the incompressibility equation is not linearly independent of the species continuity equations; if the molecular weight is discretized into $N$ components, then there will only be $N - 1$ independent values of $\phi_i$ and $v_i$ due to conservation of volume fraction and incompressibility, respectively. Arbitrarily, we may choose to let $\phi_N$ and $v_N$ be the constrained variables:

$$\phi_N = 1 - \sum_{i=1}^{N-1} \phi_i \qquad (20)$$

$$v_N \phi_N = -\sum_{i=1}^{N-1} \phi_i v_i$$

The total $x$-momentum balance (obtained by summing all species momentum balances) is given by:

$$0 = \frac{\partial}{\partial y}\left(\sigma_{xy} + Wi\eta \frac{\partial \langle u \rangle}{\partial y}\right) \qquad (21)$$

The dimensionless group $Wi = U\tau_C/H$ is the Weissenberg number of the flow for a constant moving wall velocity $U$, and $\eta = \eta_D/G\tau_C$ gives the typical scale of purely viscous forces relative to purely elastic forces in the system. Note that moving forward we have neglected the possibility of purely viscous stresses acting in the system ($\eta = 0$). If the viscous stresses arise from dissipation at sub-chain scales smaller than the tube diameter (as is often assumed) then $\eta$ should be no larger than $\eta \sim \mathcal{O}(\bar{Z}^{-3})$, and so purely viscous stresses should be negligible when (1) the melt is well entangled ($\bar{Z} \gg 1$) and (2) the deformation rate does not exceed the Rouse relaxation time of an entanglement strand ($Wi \ll \bar{Z}^3$). These two conditions are easily met in the context of our present study.

In the flow direction, the difference between any component velocity $u_i$ and the tube velocity $u_T$ is given by:

$$u_i - u_T = \left[\frac{\bar{Z}\bar{R}_g^2}{Wi}\right]\frac{\partial}{\partial y}\left(\sigma_{xy} + \left[\frac{Wi\eta}{z_m}\right]\frac{\partial \langle u \rangle}{\partial y}\right) \quad (22)$$

In the limit where $\eta = 0$ (as we consider here) the right hand side of the above equation identically vanishes and we find that all components have the same velocity in the flow direction.

To avoid solving for the pressure, we subtract the species momentum balance equations of any two components in the SRDP fluid to obtain a set of 'relative velocity equations'. Subtracting the momentum balance equations for species $i$ and $N$, we obtain:

$$z_i v_j - z_N v_N - (z_i - z_N)v_T = -\frac{\partial}{\partial y}[\mu_i - \mu_N] + \bar{Z}\alpha_{iN}\left[\frac{\partial}{\partial y}\sigma_{yy} + \sum_m \sum_n \phi_m \phi_n \frac{\partial}{\partial y}f_{mn}^{el}\right] \quad (23)$$

where $z_i = Z_i/\bar{Z}$, $\alpha_{iN} = (z_i - z_N)/\sum_m \phi_m z_m$, and $\mu_i = \delta F/\delta \phi_i$ is the chemical potential for species $i$:

$$\mu_i = \frac{1}{z_i}\left((1 + \ln(\phi_i)) - \frac{z_m}{18}\bar{R}_g^2\left[\frac{(\partial_y \phi_i)^2}{2\phi_i^2} + \frac{\partial}{\partial y}\left(\frac{1}{\phi_i}\frac{\partial \phi_i}{\partial y}\right)\right] + \right)$$
$$+ \bar{Z}N_e \sum_m \phi_m \chi_{mn} + 2\bar{Z}\sum_m \phi_m f_{im}^{el} \quad (24)$$

Finally, the constitutive equation gives:

$$\frac{\partial Q_{mn,xx}}{\partial t} + \bar{R}_g^2 v_T \frac{\partial Q_{mn,xx}}{\partial y} - 2Wi\frac{\partial u_T}{\partial y}Q_{mn,xy} = -R_{mn,xx} \quad (25)$$

$$\frac{\partial Q_{mn,yy}}{\partial t} + \bar{R}_g^2\left(v_T \frac{\partial Q_{mn,xx}}{\partial y} - 2\frac{\partial v_T}{\partial y}Q_{mn,yy}\right) = -R_{mn,yy} \quad (26)$$

$$\frac{\partial Q_{mn,zz}}{\partial t} + \bar{R}_g^2 v_T \frac{\partial Q_{mn,zz}}{\partial y} = -R_{mn,zz} \quad (27)$$

$$\frac{\partial Q_{mn,xy}}{\partial t} + \bar{R}_g^2\left(v_T \frac{\partial Q_{mn,xy}}{\partial y} - \frac{\partial v_T}{\partial y}Q_{mn,xy}\right) - Wi\frac{\partial u_T}{\partial y}Q_{mn,yy} -= -R_{mn,xy} \quad (28)$$

$$R_{mn,ij} = -\frac{1}{2}\left(\frac{1}{z_m^3} + \frac{1}{z_n^3}\right)(Q_{mn,ij} - \delta_{ij}) - 3\frac{\bar{Z}}{z_m^2}\left(1 - \frac{1}{\lambda_m}\right)\left(Q_{mn,ij} + \frac{1}{\lambda_m}(Q_{mn,ij} - \delta_{ij})\right)$$
$$- 3\frac{\bar{Z}}{z_n^2}\left(1 - \frac{1}{\lambda_n}\right)\left(Q_{mn,ij} + \frac{1}{\lambda_n}(Q_{mn,ij} - \delta_{ij})\right) \quad (29)$$

## Section A.3: Comparisons to Polymer Solutions

To date, theoretical studies of two-fluid models have been applied more frequently to polymer solutions than polymer blends. Therefore, to the reader more familiar with the polymer solution literature we take a brief aside to discuss the relationship between our blend model and a corresponding polymer solution model [14] [27] [20].

For a two-fluid model of a semidilute entangled monodipsere polymer in a small-molecule solvent, the dynamics of SID are governed by a set of six dimensionless parameters: (1) $E$, the elastic modulus relative to the osmotic modulus, (2) $\theta$, the reptation time relative to the longest Rouse time, (3) $\bar{\xi}$, the solution correlation length relative to the 'magic length', (4) $\bar{H}$ gap dimension relative to the 'magic length', (5) $Wi$, the Weissenberg number of the flow, and (6) $\varpi$, the solvent viscosity relative to the polymer viscosity. For (3) and (4), the 'magic length' is the length at which diffusion and stress relaxation occur on the same time-scale.

For both cases, the Weissenberg number, $Wi$, serves a similar purpose; with increasing $Wi$, polymers go from a relaxed isotropic configuration to one that is highly aligned, oriented with the flow, and, at high enough Wi also strongly stretched.

In polymer solutions, the timescale ratio $\theta$ effectively describes the entanglement number of the melt and so for polymer blends $\bar{Z}$ generally plays the same role. However, for bi-disperse blends of high polydispersity, only entanglements between long chains make an appreciable contribution to the stress, and $\bar{Z}\phi_L$ is a better approximant to $\theta$ [19].

In a polymer melt, the 'magic length' is trivially equal to the typical coil size – a polymer will relax its stress on the same time-scale that it is able to migrate from its tube. Thus, the dimensionless system size $\bar{H}$ plays the same role as $1/\bar{R}_g$.

In a polymer solution, the solution correlation length controls the size at which interfacial stresses suppress concentration gradients – in our model of polymer blends, we use the tube radius as the minimum interfacial length, and so the polymer solution variable $\bar{\xi}/\bar{H}$ can be compared to $\bar{R}_g/\bar{Z}^{1/2}$.

In general, the viscosity ratio $\varpi$ in polymer blends plays the same role as the viscosity ratio $\eta$ in polymer solutions. For certain special cases of high polydispersity bi-disperse blends, however, one may find more success by comparing $\varpi$ to the viscosity ratio of the component species, $\phi_S z_S^3/\phi_L^2 z_L^3$.

Finally, the parameter $E$ describes the relative scale of elastic and osmotic forces driving migration: in polymer blends, the elastic forces ($k_BT$ per entanglement) are generally $\mathcal{O}(\bar{Z})$ relative to the mixing forces ($k_BT$ per chain) but the elastic forces are only able to effect a migration to the extent that there is some rheological contrast between chains, $I_P - 1 \neq 0$. Hence, the appropriate comparison for $E$ in a homologous polymer melt is generally $E \sim \bar{Z}(I_P - 1)$ when $I_P - 1 \ll 1$. For bi-disperse blends of very high polydispersity, short chains dominate the mixing forces and the same series of scaling arguments leads to a comparison of $E \sim Z_S \phi_S$

## Section A.4: Evaluating the sign of $\sigma_0$

For linear perturbations of very long wavelength, the linear stability analysis of our governing equations is somewhat simplified. One obvious simplification is that the gradient terms inside the chemical potential can be safely ignored. More importantly, however, the growth or decay of a perturbation is limited by diffusion at long wavelengths, and so changes in polymer configuration are (to leading order) quasi-static with changes in the concentration. As a result, the 'adiabatic approximation' becomes exact [29] [20], and the configuration tensors $Q_{ij}$ become slaved to the concentration. This simplifies the linear stability analysis, since only the concentration needs to be perturbed directly: all other variables are perturbed only indirectly as a result of the perturbation to the concentration (i.e. $\delta Q_{ij} = \sum_k a_{ij,k}\delta\phi_k$).

The linear stability analysis then follows as usual. We perturb all components of the concentration independently in order to construct a linear matrix equation for the dynamics of any possible perturbation:

$$\frac{\partial}{\partial t}\begin{bmatrix}\delta\phi_1 \\ \delta\phi_2 \\ \vdots \\ \delta\phi_{N-1}\end{bmatrix} = (k\bar{R}_g)^2 \boldsymbol{M}(\bar{Z}, I_P, \ldots, Wi)\begin{bmatrix}\delta\phi_1 \\ \delta\phi_2 \\ \vdots \\ \delta\phi_{N-1}\end{bmatrix} \qquad (30)$$

The largest eigenvalue of the matrix $\boldsymbol{M}$ provides the value of $\sigma_0$, and the corresponding eigenvector describes the fastest growing (or slowest decaying) perturbation to the molecular weight distribution.